\documentclass[useAMS,usenatbib,twocolumn,scrartcl]{mnras}
\usepackage{epsfig,amssymb}
\usepackage{graphicx,amsmath,multicol}
\usepackage{multirow}
\usepackage{caption}
\usepackage{subcaption}
\usepackage{placeins}
\usepackage{physics}

\usepackage{hyperref}
\hypersetup{
     colorlinks = true,
     linkcolor = blue,
     urlcolor = red,
     citecolor = blue
}

\usepackage{footnote}
\usepackage{tablefootnote}
\makesavenoteenv{tabular}
\makesavenoteenv{table}
\usepackage{supertabular}

\usepackage{enumerate}

\usepackage[utf8]{inputenc}
\usepackage{setspace}
\usepackage{graphicx}
\usepackage{mathtools}
\usepackage{grffile}
\usepackage{amsmath}
\usepackage{longtable}
\usepackage{enumitem}

\usepackage{graphics}
\usepackage{verbatim}
\usepackage[usenames]{color}
\usepackage{float}

\def\apj{{ ApJ}}
\def\apjl{{ ApJL}}

\def\aap{{ A\&A}}

\def\mnras{{ MNRAS}}
\def\araa{{ ARA\&A}}

\def\sci{{ Science}}
\def\nat {{ Nature}}

\def\pasa{{ PASA}}

\title[FRB properties from observables]{Constraining FRB progenitors from flux distribution} 
\author[Bhattacharya]
{Mukul Bhattacharya\\ 
Department of Physics, University of Texas at Austin, Austin, TX 78712, USA} 

\begin{document}

\date{Accepted . Received ; in original form }

\pagerange{\pageref{firstpage}--\pageref{lastpage}} \pubyear{2019}

\maketitle

\label{firstpage}

\begin{abstract}
We present a generic formalism to constrain the cosmic rate density history and the source luminosity function of FRB progenitors from the statistical properties of the apparent flux density of non-repeating FRBs detected with Parkes. We include the pulse multipath propagation effects to evaluate the flux density distribution for a generalised spatial density model and luminosity function. We perform simulations to investigate the effects of the telescope beam pattern and temporal resolution on the observed flux density. We find that the FRB progenitors are likely to be younger stars with a relatively flat energy spectrum and host galaxy DM contribution similar to the MW. Our analysis can be extended to larger FRB samples detected with multiple surveys to place stronger constraints on the FRB progenitor properties. 
\end{abstract}

\begin{keywords}
radio continuum: transients - cosmology: observations - scattering - turbulence - ISM: general \vspace{-0.3cm}
\end{keywords}

\section{Introduction}
\label{Intro}
Fast radio bursts (FRBs) are bright transients of unknown physical origin that last for short durations of few milliseconds and have been detected at radio frequencies between 400 MHz and 8 GHz \citep{Lorimer07,Thornton13,Petroff16}. Their large dispersion measures (DMs) are found to be well in excess of the Galactic interstellar medium (ISM) contribution, which indicates an extragalactic origin of these bursts. Currently, about 70 distinct FRB events have already been reported \citep{Petroff16}, with FRB 121102 \citep{Scholz16,Spitler16} and FRB 180814.J0422+73 \citep{CHIME19} being the only repeating sources in that sample. The localization of the repeating FRB 121102 within a star-forming region in a dwarf galaxy at redshift $z=0.19$ confirmed the cosmological origin of this source \citep{Chatterjee17,Marcote17,Tendulkar17}. 
As more FRB sources get localized in the near future, FRBs can  be potentially used as cosmological probes to study the baryonic distribution within the IGM as well as to constrain the cosmological parameters in our Universe \citep{Gao14,Zheng14}.

Although several progenitor models involving both cataclysmic and non-cataclysmic scenarios have already been proposed for FRBs (see \citealt{Platts18}, for a recent review), the nature of FRBs and their sources still remains a mystery. This is primarily due to the sparse spatial localisation of several arcminutes for most of the current radio surveys, which makes the identification of the FRB host galaxy and its association with other electromagnetic counterparts challenging. However, the distributions of FRB observables such as the flux density and fluence helps us to statistically constrain the properties of the FRB progenitors as they are linked to the source luminosity function as well as the evolutionary history of the cosmic rate density \citep{Bera16,Caleb16,Oppermann16,Vedantham16,ME18,Niino18,Bhattacharya19}.   

The distribution of the observed flux density is mainly affected by the pulse temporal broadening due to multipath propagation and the finite temporal resolution of the detection instrument. In this work, we investigate how the statistical properties of the apparent flux density can be used to constrain the luminosity and spatial density distributions of the FRB progenitors for events detected specifically with Parkes. We consider the effects of the telescope beam shape and temporal resolution on the observed flux distribution in addition to the pulse propagation effects.  
Due to the rapidly evolving nature of this field, we only consider the FRBs published until February 2019 with resolved intrinsic width and total $DM \geq 500\ {\rm pc\ cm^{-3}}$ for our analysis here (see Table 1 of \citealt{Bhattacharya19} for the data sample). We assume fiducial values for cosmological parameters with $H_{0} = 68 {\rm \ km\ s^{-1}}{\rm Mpc^{-1}}$, $\Omega_{m} = 0.27$ and $\Omega_{\Lambda} = 0.73$ \citep{Planck14}.

This \emph{Letter} is organized as follows. In Section 2, we estimate the FRB distances and flux densities assuming specific host galaxy properties and scattering model for pulse temporal broadening. In Section 3, we obtain the flux density distribution for a given FRB spatial density and luminosity distribution to compare it with the current observations. We then perform Monte Carlo (MC) simulations to study the effect of telescope observing biases, source energy density function and host galaxy properties on the observed flux distribution in Section 4. We conclude with a summary of our results in Section 5. 

\section{FRB distance and flux estimates}
\label{sec2}
The inferred FRB distances are based on the uncertain host galaxy properties and the source location inside it, with a larger uncertainty in $z$ expected for a larger host galaxy DM contribution ($DM_{\rm host}$) to the total DM ($DM_{\rm tot}$). In order to minimize the uncertainty from $DM_{\rm host}$, we place a lower $DM_{\rm tot}$ cutoff on the observed sample considered in this study. The total DM for a given FRB line of sight is
\begin{equation}
DM_{tot} = DM_{\rm MW} + \frac{DM_{\rm host}}{(1+z)} + DM_{\rm IGM,0} \int_{0}^{z}\frac{(1+z^{\prime})dz^{\prime}}{\sqrt{(1+z^{\prime})^3 + 2.7}}
\label{DMtot}
\end{equation}
where $DM_{\rm MW}$ is the Milky Way (MW) ISM contribution obtained from the NE2001 model \citep{CL02} and the IGM DM contribution ($DM_{\rm IGM}$) is given by the integral over $z$ with $DM_{\rm IGM,0} = 1294.9\ {\rm pc\ cm^{-3}}$. We have assumed the baryonic mass fraction in the IGM $f_{IGM} = 0.83$, free electron number density $n_e(z) = 2.1\times10^{-7} (1+z)^3 \ {\rm cm^{-3}}$ and the ionization fraction $x(z) \approx 7/8$ \citep{Ioka03,Inoue04,DZ14}. As the type of the host galaxy, location of the FRB source within its galaxy and our viewing angle relative to the host galaxy are all fairly uncertain, we assume a fixed contribution $DM_{\rm host} \approx 100\ {\rm pc\ cm^{-3}}$ that is comparable to the typical $DM_{\rm MW}$ contribution for most lines of sight. The inferred redshift is then obtained directly from equation (\ref{DMtot}) for a given burst and the corresponding comoving distance to the source is $D(z) = (8.49\ {\rm Gpc}) \int_{0}^{z} [(1+z^{\prime})^3 + 2.7]^{-0.5} dz^{\prime}$. For a typical uncertainty $\Delta DM_{\rm host} \sim DM_{\rm host} \approx 100\ {\rm pc\ cm^{-3}}$ and $DM_{\rm IGM} \approx 750z\ {\rm pc\ cm^{-3}}$, the corresponding error in the inferred redshift is $\Delta z \approx 0.2/(1+z)$ assuming $\Delta z \sim z \sim 1$ for most FRBs in our sample.

The intrinsic width of a FRB pulse can be written in terms of the observed width $w_{\rm obs}$ and other width components as $w_{\rm int}^{2} = [w_{\rm obs}^{2} - (w_{\rm DM}^2 + w_{\rm samp}^2 + w_{\rm sc}^2)]/(1+z)^2$. Here, $w_{\rm DM}$ is the dispersive smearing across frequency channels, $w_{\rm samp}$ is the observation sampling time and $w_{\rm sc}^{2} = w_{\rm IGM}^{2} + w_{\rm ISM,MW}^{2} + (1+z)^{2} w_{\rm ISM,host}^{2}$ is the pulse scatter broadening across the diffused IGM and ISM components. The scatter broadening in the host galaxy/MW ISM is given by \citep{Krishnakumar15}
\begin{eqnarray}
\begin{aligned}
\label{wISMhostMW}
&w_{\rm ISM} = w_{\rm ISM,0} F (1.0+1.94\times10^{-3} DM_{\rm ISM}^{2.0})\frac{DM_{\rm ISM}^{2.2}}{\nu_{0,GHz}^{4.4}}
\end{aligned}
\end{eqnarray}
where $DM_{\rm ISM} = DM_{\rm host/MW}$ is the respective ISM DM component, $w_{\rm ISM,0} = 4.1\times10^{-8}\ {\rm ms}$, $\nu_{0,GHz} = \nu_{0}/10^3$ is the central frequency in GHz and $F = 4f(1-f)$ is the geometrical lever-arm factor with $f$ = 25 kpc/$[(1+z)D(z)]$ \citep{Williamson72,Vandenberg76}. The pulse broadening due to IGM turbulence is \citep{MK13}
\begin{eqnarray}
w_{IGM}(z) = \frac{k_{IGM}}{\nu_{0,GHz}^{4}Z_{L}}\int^{z}_{0} dz^{\prime} d(z^{\prime})\int^{z}_{0} dz^{\prime} d(z^{\prime}) (1+z^{\prime})^{3}
\label{wIGM2}
\end{eqnarray}
where $k_{IGM} = 2.94\times10^{12}\ {\rm ms\ MHz^{4}}$ is a normalisation factor, $Z_{L} = (1+z)^{2}\left[(1+z) - \sqrt{z(1+z)}\right]^{-1}$ and $d(z^{\prime}) = [\Omega_{m}(1+z^{\prime})^3 + \Omega_{\Lambda}]^{-1/2}$. As the value of $w_{\rm sc}$ from equations (\ref{wISMhostMW}) and (\ref{wIGM2}) is significantly smaller compared to $w_{\rm obs}$ and $w_{\rm DM}$ (see \citealt{Bhattacharya19}), we have $\Delta w_{\rm int}/w_{\rm int} = 0.2/(1+z)^{2}$. We find the best-fit cumulative $w_{\rm int}$ distribution to be: $N(>w_{\rm int}) = 25.24\ {\rm exp}(-w_{\rm int}/2.092\ {\rm ms})$.
Although the value of $k_{\rm IGM}$ is fixed using the width parameters of a particular FRB, $w_{\rm int}$ has a weak dependence on the constant $k_{\rm IGM}$ as $w_{\rm sc} \lesssim 10^{-2} w_{\rm int}$. 

\begin{figure}
  \begin{subfigure}[tp]{1.02\linewidth}
    \centering
    \includegraphics[height=0.67\linewidth,width=0.93\linewidth]{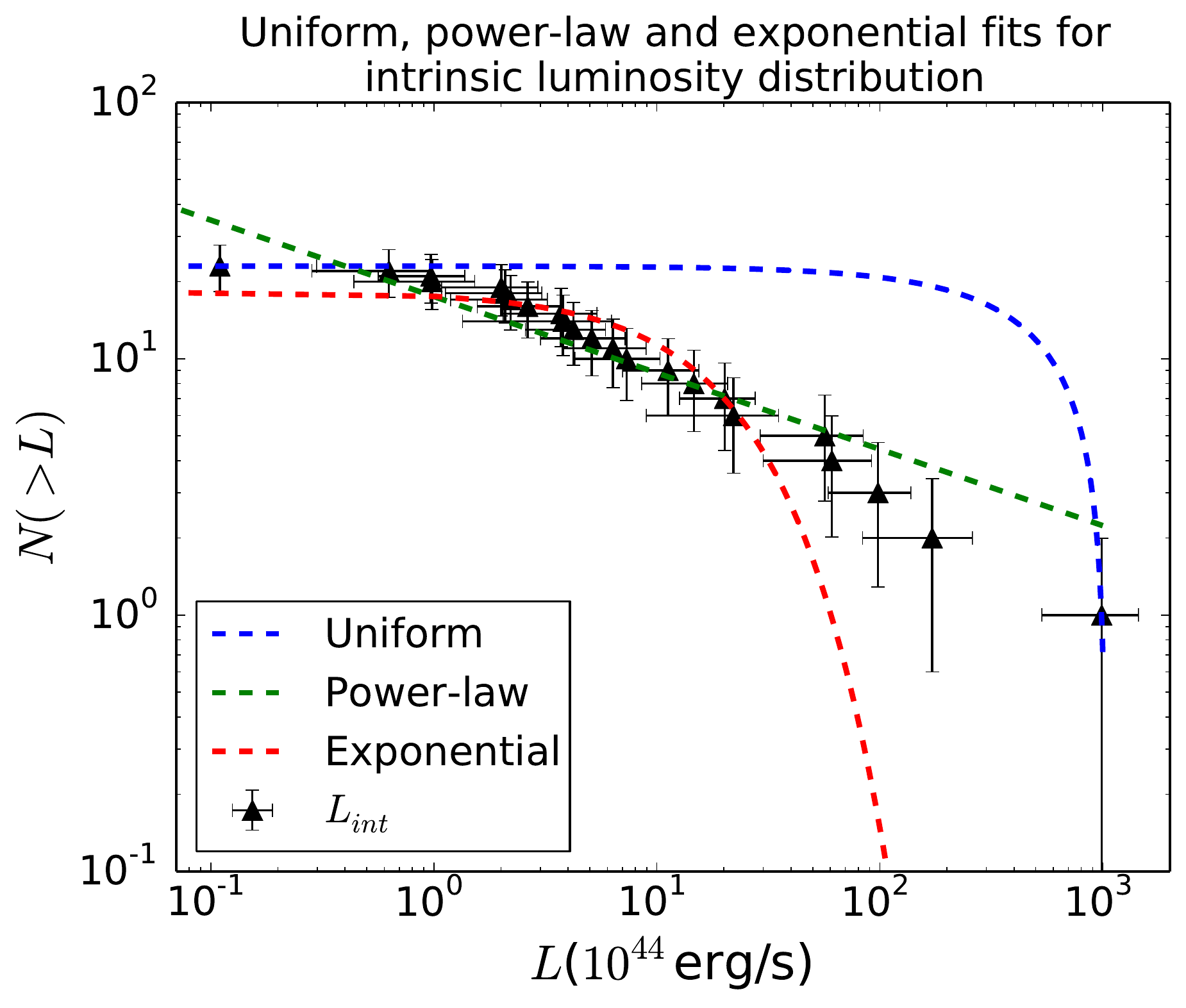} 
  \end{subfigure}  	
	\caption[Small multiples]{\small \emph{Chi-squared fits for the cumulative distribution of the inferred luminosities:} $L$ values are obtained from equation (\ref{EdistplLE}) and vary from $1.1\times10^{43}\ {\rm erg/s}$ to $1.0\times10^{47}\ {\rm erg/s}$ with the best fit: $23.0 - 0.0223\ L$ uniform, $17.489\ L^{-0.298}$ power-law and $18.395\ {\rm exp}(-L/20.741\times10^{44}\ {\rm erg/s})$ exponential distributions. \vspace{-0.7cm}
}%
\label{fig1}
\end{figure}

\begin{table*}
\begin{center}
\caption{Polynomial approximations for $N(>S_{\rm peak})$ distributions obtained from equation (\ref{NS}) for various distributions of FRB source luminosity function $g(L)$ and spatial density $\rho(z)$. We define $\xi = L_{\rm min}/L_{\rm mid}$, $R_{\rm min} = (1.114\ {\rm Gpc})S_{\rm peak}^{-1/2}$, $R_{\rm mid} = (7.598\ {\rm Gpc})S_{\rm peak}^{-1/2}$, $F_{\rm SFH,PL}(x) = 2.03 + 4.96 x - 0.874 x^{2} + 0.131 x^{3} - 0.00663 x^{4}$, $F_{\rm SMD,PL}(x) = 6.54 + 0.0155 x + 0.950 x^{2} + 0.143 x^{3} - 0.0517 x^{4}$.
}
\label{Table1}
\bgroup
\def\arraystretch{0.8}
\begin{tabular}{| c | c | c |}
\hline
\hline
\centering
$\rho(z)$ & $g(L) = 1/(L_{\rm mid} - L_{\rm min})$ & $g(L) \propto L^{-1.298}$ \\ \hline \hline
$\rho_{\rm NE}(z)$ & $\propto S_{\rm peak}^{-3/2}(1 - \xi^{5/2})$ & $\propto S_{\rm peak}^{-0.298}(1 - 0.00406/S_{\rm peak}^{1.202})$ \\ \hline
$\rho_{\rm SFH}(z)$ & $\propto [-0.606(1 - \xi) + 0.668 R_{\rm mid}(1 - \xi^{3/2}) - 0.307 R_{\rm mid}^{2}(1 - \xi^{2})$ & $\propto [S_{\rm peak}^{-0.298} F_{\rm SFH,PL}(R_{\rm max})-3.92 F_{\rm SFH,PL}(R_{\rm min})] $ \\ 
& $+ 0.0632 R_{\rm mid}^{3}(1 - \xi^{5/2}) - 0.00376 R_{\rm mid}^{4}(1 - \xi^{3})]$ & \\ \hline
$\rho_{\rm SMD}(z)$ & $\propto [-1.95(1 - \xi) + 0.00208 R_{\rm mid}(1 - \xi^{3/2}) + 0.334 R_{\rm mid}^{2}(1 - \xi^{2})$ & $\propto [S_{\rm peak}^{-0.298} F_{\rm SMD,PL}(R_{\rm max})-3.92 F_{\rm SMD,PL}(R_{\rm min})] $ \\ 
& $+ 0.0688 R_{\rm mid}^{3}(1 - \xi^{5/2}) - 0.0293 R_{\rm mid}^{4}(1 - \xi^{3})]$ & \\ \hline
\hline \vspace{-0.4cm}
\end{tabular}
\egroup
\end{center}
\end{table*}

The flux density is reduced due to the pulse broadening from multipath propagation and can be written in terms of the observable fluence $\mathcal{F}_{\rm obs}$ as $S_{\rm peak} = \mathcal{F}_{\rm obs}/w_{\rm int}$. Moreover, for a Gaussian telescope beam profile the observed flux density is further reduced with $S_{\rm peak,obs} \approx S_{\rm peak}\ {\rm exp}(-r^{\prime 2}/r_{\rm beam}^{2})$, where $r^{\prime}$ is the radial distance from the center of the beam with radius $r_{\rm beam}$. The bolometric luminosity is obtained for a power-law FRB energy distribution to be \citep{Lorimer13}
\begin{equation}
L = \frac{4\pi D^{2}(z) (\nu_{\rm max}^{\prime \alpha+1} - \nu_{\rm min}^{\prime \alpha+1}) (\nu_2 - \nu_1)}{(1+z)^{\alpha-1} (\nu_{2}^{\alpha+1} - \nu_{1}^{\alpha+1})} S_{\rm peak}  
\label{EdistplLE}
\end{equation}
where $\alpha$ is the spectral index of the energy distribution, $(\nu_{min}^{\prime},\nu_{max}^{\prime})$ is frequency range for source emission in the comoving frame and $(\nu_{1},\nu_{2})$ is the observing band frequency range. As the emission spectral indices are poorly constrained from the current observations, we assume $\alpha \approx 0$, $\nu_{min}^{\prime} = 600\ {\rm MHz}$ and $\nu_{max}^{\prime} = 8\ {\rm GHz}$ for our analysis here. As $D(z) \propto z$, the relative uncertainty in the inferred luminosity is $\Delta L/L = \sqrt{(\Delta S_{\rm peak}/S_{\rm peak})^{2} + 4(\Delta z/z)^{2}}$.

Figure \ref{fig1} shows the chi-squared fits for the uniform, power-law and exponential cumulative distributions of the inferred FRB luminosities from equation (\ref{EdistplLE}). We find that the luminosity varies considerably by almost four orders of magnitude from $\sim 10^{43}\ {\rm erg/s}$ to $\sim 10^{47}\ {\rm erg/s}$ for our sample. We include the uncertainties $\Delta L$ as the x-error bars along with the Poisson fluctuations as errors in the y-coordinate to obtain the chi-squared fits. While we find that both the power-law (PL) distribution $\propto L^{-0.298}$ and the exponential distribution $\propto {\rm exp}(-L/L_{\rm c})$ with cutoff $L_{\rm c} \approx 20.741\ {\rm erg/s}$ fit the inferred luminosity fairly well, the former explains the relative over-abundance of non-repeating FRBs with very large inferred luminosities $\gtrsim 10^{46}\ {\rm erg/s}$ better.

\section{Observed flux distribution}
\label{sec3}
For a population of FRB sources distributed within a distance $R_{\rm min}$ to $R_{\rm max}$, the number of sources with luminosity $L$ having peak flux density larger than some $S_{\rm peak}$ are
\begin{equation}
\label{Eqn1}
\frac{dN}{dL} = \left\{
\begin{array}{ll}
{\displaystyle \int_{R_{min}}^{D_L}} {\displaystyle \frac{n(z,L)}{1+z}} 4\pi D^{2} dD, \ R_{\rm min}^{2} < L/4\pi S_{\rm peak} < R_{\rm max}^{2} \vspace{0.1cm}\\ 
{\displaystyle \int_{R_{min}}^{R_{max}}} {\displaystyle \frac{n(z,L)}{1+z}} 4\pi D^{2}dD, \ L/4\pi S_{\rm peak} > R_{\rm max}^{2} 
\end{array}
\right. 
\end{equation}
where $D_L = \sqrt{L/4\pi S_{\rm peak}}$ and we assume that $n(z,L) = \rho(z)g(L)$ is independent of $w_{\rm int}$. Here $g(L)$ is the luminosity distribution of the FRB source and $\rho(z)$ is the spatial density distribution of the FRB progenitors. As FRBs with relatively small $DM_{\rm tot} \sim 100\ {\rm pc\ cm^{-3}}$ have already been reported, here we set $R_{\rm min} = 0$ and $R_{\rm max} \approx 11\ {\rm Gpc}$ such that $R_{\rm max} \geq (L/4\pi S_{\rm peak})^{1/2}$ holds for all the FRBs in our data sample. This further gives the source count to be
\begin{eqnarray}
N(>S_{\rm peak}) = 4\pi {\displaystyle \int_0^{4\pi R_{\rm max}^{2} S_{\rm peak}}} g(L)dL \int_0^{D_L} \frac{\rho(z)}{1+z} D^2 dD
\label{NS}
\end{eqnarray}
where $N(>S_{\rm peak})$ is directly determined from the observations with $g(L)$ and $\rho(z)$ decided by the nature of the FRB progenitor.

\begin{figure}
   \begin{subfigure}[tp]{1.02\linewidth}
    \centering
    \includegraphics[height=0.67\linewidth,width=0.93\linewidth]{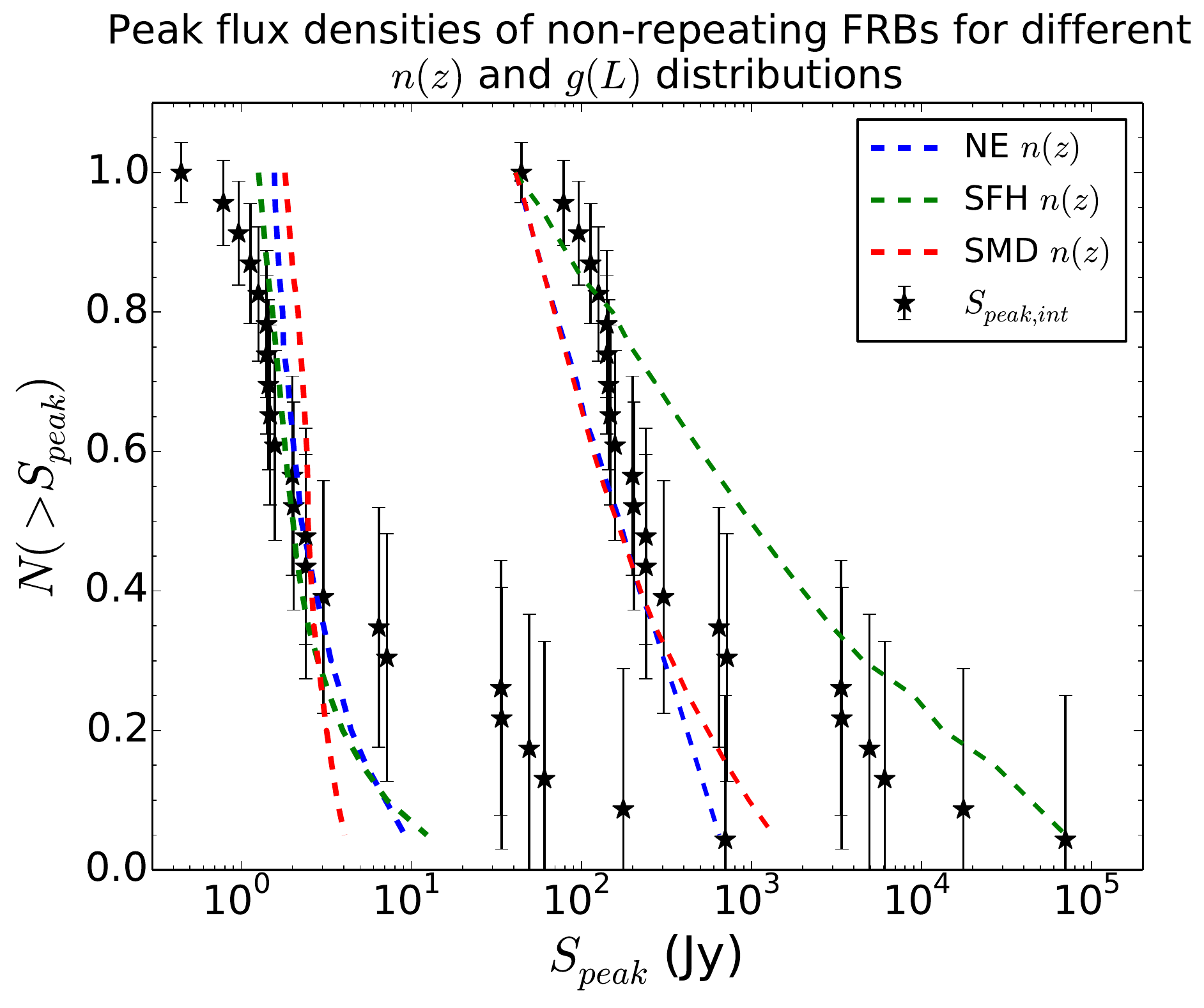} 
  \end{subfigure}
	\caption[Small multiples]{\small \emph{Comparison of intrinsic $S_{\rm peak} = \mathcal{F}_{\rm obs}/w_{\rm int}$ distribution with that obtained for different $g(L)$ and $\rho(z)$ models:} $N(>S_{\rm peak})$ distributions obtained from equation (\ref{NS}) for uniform/power-law $g(L)$ and NE/SFH/SMD $\rho(z)$ are shown here. The $S_{\rm peak}$ values for the bursts are scaled up from their actual values by a factor of 100 for the power-law $g(L)$ to avoid overlap.  \vspace{-0.6cm} 
}%
	\label{fig2}
\end{figure}

We consider uniform $g(L) \propto 1/(L_{\rm mid}-L_{\rm min})$ and power-law $g(L) \propto L^{-1.298}$ distributions, where $L_{\rm min} = 1.1\times10^{43}\ {\rm erg/s}$ and $L_{\rm mid} = 5.1\times10^{44}\ {\rm erg/s}$ are the minimum and median inferred luminosities from our sample. For the spatial density $\rho(z)$, we consider three different models: (a) non-evolving population $\rho_{\rm NE}(z)$ of FRB progenitors, (b) spatial density tracking the star formation history (SFH) $\rho_{\rm SFH}(z)$, and (c) spatial density tracking the stellar mass density (SMD) $\rho_{\rm SMD}(z)$. We use the formulations of cosmic SFH and SMD given by \citet{MD14},
\begin{eqnarray}
\rho_{\rm SFH}(z) = \psi(z) = \rho_{\rm SFH,0}\frac{(1+z)^{2.7}}{1+[(1+z)/2.9]^{5.6}} \\
\rho_{\rm SMD}(z) = \rho_{\rm SMD,0}\int_{z}^{\infty}\psi(z^{\prime})\frac{d(z^{\prime})dz^{\prime}}{H_0(1+z^{\prime})}
\end{eqnarray}
where $\rho_{\rm SFH,0} = 0.015\ M_{\odot}\ {\rm year}^{-1}\ {\rm Mpc}^{-3}$ and $\rho_{\rm SMD,0} = 0.73$ are the normalisation constants. While $\rho(z)$ is expected to follow $\rho_{\rm SFH}(z)$ if FRBs arise from relatively young population of stars, the spatial density should trace $\rho_{\rm SMD}(z)$ if FRB progenitors were to be older stars.

Table \ref{Table1} lists the closest polynomial approximations for the cumulative flux distributions obtained for the two source luminosity functions and the three FRB spatial density models considered here. In Figure \ref{fig2}, we show the comparison of the intrinsic $S_{\rm peak} = \mathcal{F}_{\rm obs}/w_{\rm int}$ obtained in Section \ref{sec2} with that computed from equation (\ref{NS}) for the different $g(L)$ and $\rho(z)$ models used here. The p-values from the Kolmogorov Smirnov (KS) test comparison between the distributions are listed in Table \ref{Table2}. We find that the distribution of intrinsic $S_{\rm peak}$ is better explained by a young population of FRB progenitors with $\rho(z) \propto \rho_{\rm SFH}$, especially for a uniform $g(L)$. While the $\rho_{\rm NE}$ and $\rho_{\rm SMD}$ spatial densities can be ruled out for uniform $g(L)$, all three $\rho(z)$ distributions explain the flux density values in case of a power-law $g(L)$ fairly well. \vspace{-0.15cm}

\section{Observing biases}
\label{sec4}

\begin{figure}
   \begin{subfigure}[tp]{1.02\linewidth}
    \centering
    \includegraphics[height=0.67\linewidth,width=0.93\linewidth]{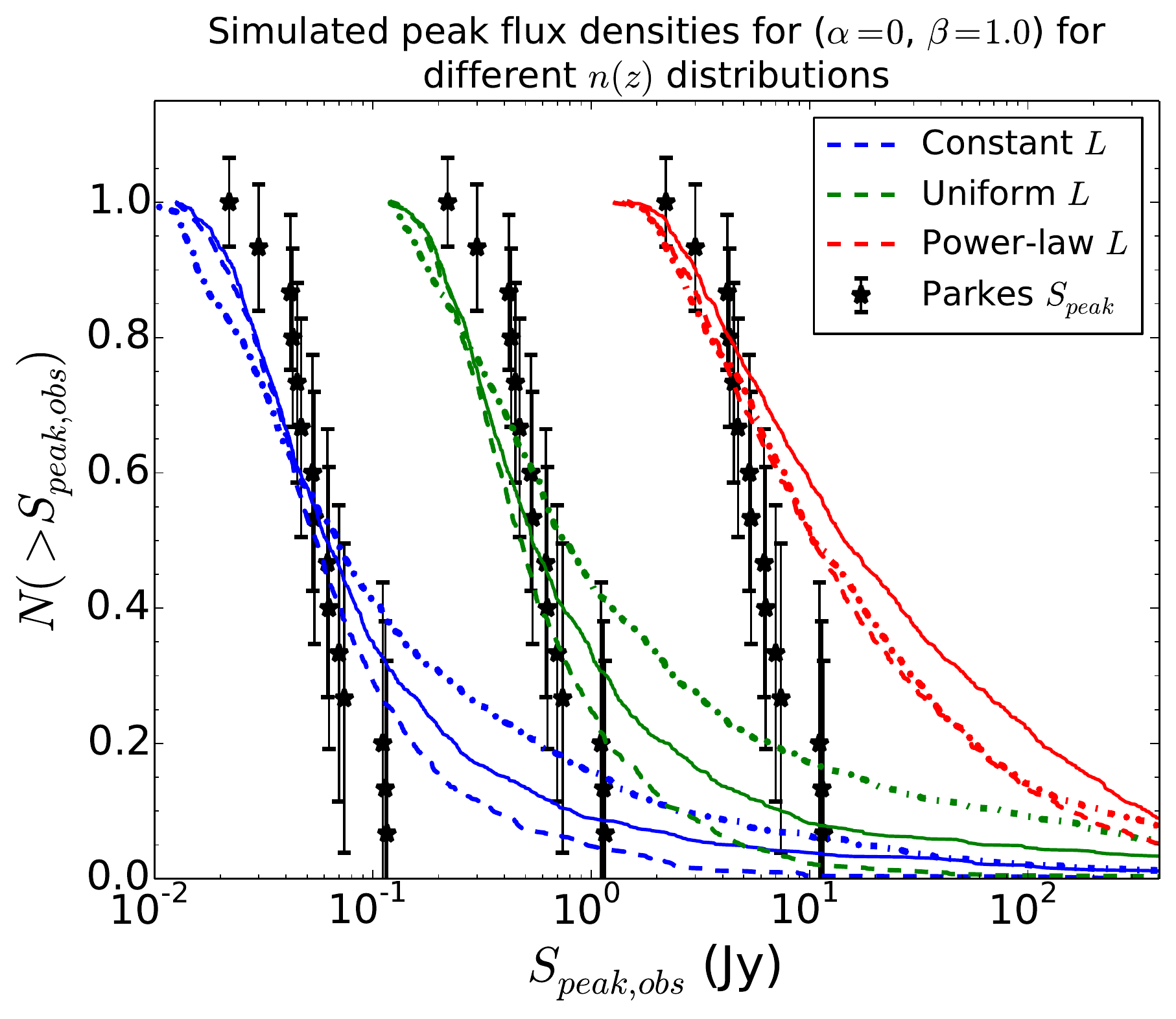} 
  \end{subfigure}
	\caption[Small multiples]{\small \emph{Comparison of observed Parkes $S_{\rm peak}$ with simulated $S_{\rm peak}$ distributions for different $g(L)$ and $\rho(z)$ models:} Simulated $N(>S_{\rm peak,obs})$ distributions for constant/uniform/power-law $g(L)$ and NE/SFH/SMD $\rho(z)$ are shown for $\alpha=0$ and $\beta=1.0$. We rescale the $S_{\rm peak,obs}$ values by a factor of 0.1/1/10 for the constant/uniform/power-law $g(L)$ in order to avoid overlap. The solid/dotted/dot-dashed lines for each $g(L)$ denote the NE/SFH/SMD $\rho(z)$ distribution. 
}%
	\label{fig3}
\end{figure}

\begin{table*}
\begin{center}
\caption{KS test p-values obtained from the comparison of $N(>S_{\rm peak})$ distributions in Figure \ref{fig2} (from equation \ref{NS}) and Figure \ref{fig3} (from MC simulations) with the observed FRB population. We consider constant, uniform and power-law luminosity distributions along with NE (SFH) [SMD] spatial density models. 
}
\label{Table2}
\bgroup
\def\arraystretch{0.8}
\begin{tabular}{| c | c | c | c|}
\hline
\hline
\centering
Case & $g(L) = \delta(L - L_{\rm mid})$ & $g(L) = 1/(L_{\rm mid} - L_{\rm min})$ & $g(L) \propto L^{-1.298}$ \\ \hline \hline
Equation \ref{NS} & & 0.055 (0.402) [0.024] & 0.227 (0.214) [0.306] \\ \hline
$(\alpha=0,\beta=1)$ & 0.084 (0.247) [0.019] & 0.111 (0.225) [0.009] & $3.05\times10^{-4}$ ($3.05\times10^{-3}$) [$2.03\times10^{-3}$] \\ \hline
$(\alpha=-1.4,\beta=1)$ & 0.040 (0.070) [$2.49\times10^{-4}$] & 0.025 (0.127) [$2.41\times10^{-4}$] & $3.48\times10^{-4}$ ($2.49\times10^{-3}$) [$2.03\times10^{-3}$] \\ \hline
$(\alpha=0,\beta=10)$ & 0.127 (0.156) [0.015] & 0.159 (0.162) [0.008] & $4.69\times10^{-4}$ ($1.41\times10^{-3}$) [$3.05\times10^{-4}$] \\ \hline
$(\alpha=-1.4,\beta=10)$ & 0.034 (0.125) [$3.72\times10^{-4}$] & 0.047 (0.156) [$3.48\times10^{-4}$] & $1.77\times10^{-4}$ ($1.64\times10^{-3}$) [$3.14\times10^{-3}$] \\ \hline
\hline
\end{tabular}
\egroup
\end{center}
\end{table*}

We evaluated $N(>S_{\rm peak})$ for given $g(L)$ and $\rho(z)$ models in Section \ref{sec3}, and also obtained the width distribution $N(>w_{\rm int}) = 25.24\ {\rm exp}(-w_{\rm int}/2.092\ {\rm ms})$ in Section \ref{sec2}. However, the pulse width distribution is directly affected by the temporal resolution of the telescope as a coarse time resolution makes it harder to detect a pulse with smaller $w_{\rm obs}$ due to the instrumental noise. Furthermore, there is an observing bias against bursts that are smeared over larger $w_{\rm obs}$ and/or have larger $w_{\rm int}$, as the instrument sensitivity decreases gradually with increasing $w_{\rm obs}$. In addition to the instrument temporal resolution, the observed flux distribution is also affected by the beam shape of the telescope used for the event detection. To include the effect of these observing biases on $N(>S_{\rm peak})$, we perform MC simulations to obtain the flux distribution (see Section 3.2 of \citealt{Bhattacharya19} for a detailed code algorithm).

From the known $N(>w_{\rm int})$, $g(L)$ and $\rho(z)$ distributions, we generate a population of 1000 FRBs that can be detected at the Parkes multibeam (MB) receiver with a signal-to-noise ratio $S/N \geq 9$. The Parkes MB receiver has 13 beams with beam radii $r_{beam} = 7.0^{\prime}\ (7.05^{\prime})\ [7.25^{\prime}]$ and beam center gains $G_{beam} = 0.731\ (0.690)\ [0.581]\ {\rm K\ Jy^{-1}}$ for beam 1\ (2-7)\ [8-13] (see \citealt{Stav96} for the system parameters). As the FRB source location within its host galaxy is highly uncertain, we assume for simplicity that all the detected bursts are located at the position of the Solar system. We estimate the host galaxy DM contribution as $DM_{\rm host} = \beta DM_{\rm NE2001}$, where $\beta$ is the scaling factor related to the host galaxy size compared to the MW and $DM_{\rm NE2001}$ is predicted by the NE2001 model \citep{CL02}. The assumption about the location of the FRB source will not affect our analysis here qualitatively as $DM_{\rm tot} \gg DM_{\rm host}$ for most of the reported bursts. 

In Figure \ref{fig3}, we show the comparison of the observed $S_{\rm peak}$ at Parkes with that obtained from the simulations for different $g(L)$ and $\rho(z)$ models. We perform these simulations for constant, uniform and power-law $g(L)$ along with NE, SFH and SMD $\rho(z)$ distributions. We also vary the energy density spectral index $\alpha=0,-1.4$ and the $DM_{\rm host}$ parameter $\beta=1,10$. The best-fit value of $\alpha \approx -1.5$ was recently obtained by \citet{Macquart2019} from the spectra of 23 FRBs detected with ASKAP \citep{Bannister17,Shannon18}. We list the p-values obtained from the KS test comparison for all these cases in Table \ref{Table2}. We perform all KS tests under the null hypothesis that the two samples were drawn from the same distribution unless the p-value $<$ 0.05. 

We find that FRBs most likely do not originate from older stars as $\rho(z) \propto \rho_{\rm SMD}$ is unfavored by the current Parkes observations for all $g(L)$ models and $(\alpha,\beta)$ combinations. Moreover, power-law $g(L) \propto L^{-1.298}$ over-estimates the occurrence of brighter events for all FRB spatial density distributions. The FRB source luminosity distribution is better modelled with a sharp cutoff around $L_{\rm mid} \approx 5\times10^{44}\ {\rm erg/s}$. For all $g(L)$ distributions and $(\alpha,\beta)$ combinations, we find that the FRB progenitors are most likely to be younger stars with population density history tracing the cosmic SFH as the likelihood of $\rho(z) \propto \rho_{\rm SFH}$ is found to be larger compared to $\rho(z) \propto \rho_{\rm NE}$. Lastly, while $\rho(z) \propto \rho_{\rm NE}$ with $\alpha = 0$ and $\beta \sim 1-10$ is a likely scenario, $\rho(z) \propto \rho_{\rm SFH}$ is the most favoured possibility from the current observations for $\alpha = 0$ and $\beta \sim 1$. Most events are therefore expected to arise from young stars with a relatively flat energy density distribution and a host galaxy DM contribution similar to that of the MW.

\section{Summary and conclusions}
\label{sec5}
\begin{figure}
   \begin{subfigure}[tp]{1.02\linewidth}
    \centering
    \includegraphics[height=0.7\linewidth,width=0.95\linewidth]{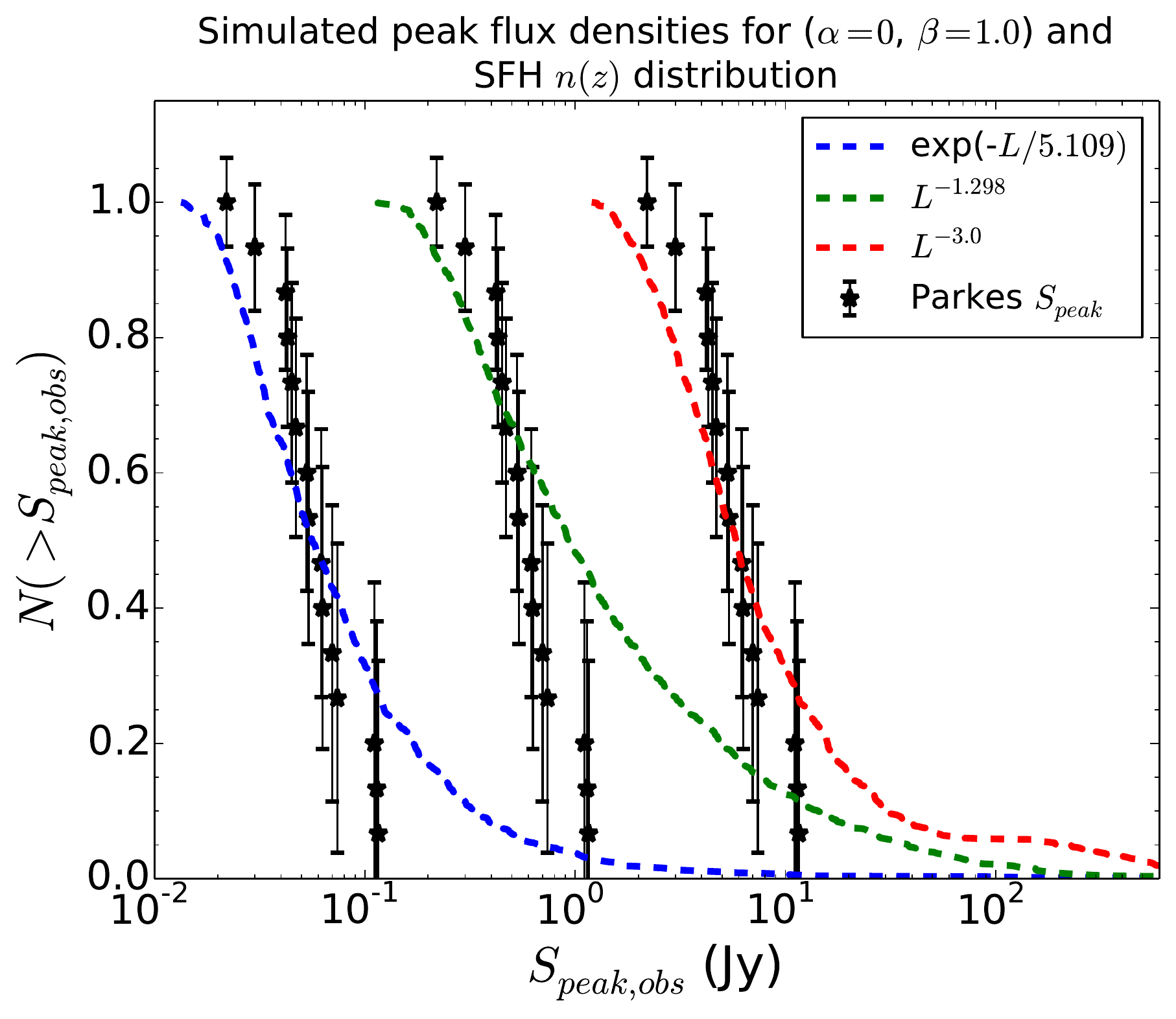} 
  \end{subfigure}
  	\caption[Small multiples]{\small \emph{Comparison of observed Parkes $S_{\rm peak}$ with simulated distributions for different $g(L)$ functions:} Simulated $N(>S_{\rm peak})$ distributions for exponential and power-law $g(L)$ with indices -1.298 and -3.0 are shown for $\alpha=0$, $\beta=1.0$ and SFH $n(z)$. We rescale the flux values for ${\rm exp}(-L/5.1\times10^{44}\ {\rm erg/s})$/$L^{-1.298}$/$L^{-3.0}$ by a factor of 0.1/1/10 to avoid overlap. The corresponding p-values are 0.196, $3.05\times10^{-3}$ and 0.213. \vspace{-0.6cm}
}%
	\label{fig4}
\end{figure}
In this \emph{Letter}, we have presented a method to constrain the source luminosity function and spatial density distribution of the FRB progenitors from the statistical properties of the observable flux density. As the sample of the reported FRBs is rapidly growing and largely heterogenous, we restrict our analysis to the Parkes FRBs that were published until February 2019 with $DM_{\rm tot} \geq 500\ {\rm pc\ cm^{-3}}$ and have resolved intrinsic widths.
We apply a lower $DM_{\rm tot}$ cutoff to minimize the errors in the distance estimates and subsequently the inferred luminosities that are based on the assumptions about the host galaxy properties and the source location inside it. Here we consider $\rho(z)$ corresponding to a non-evolving population $\rho_{\rm NE}$/young stellar population tracking $\rho_{\rm SFH}$/older stellar population tracking $\rho_{\rm SMD}$ along with constant/uniform/power-law $g(L)$ distributions. 

Assuming scattering model for pulse temporal broadening from multipath propagation and a fixed $DM_{\rm host}$ contribution, we derived $N(>S_{\rm peak})$ for a FRB population with given spatial density and luminosity function. We found that the intrinsic $N(>S_{\rm peak})$ distribution for the FRBs observed with Parkes is likely due to a population density of young stars $\propto \rho_{\rm SFH}$ and luminosity function with a sharp cutoff around $L_{\rm mid} \sim 5.1\times10^{44}\ {\rm erg/s}$. While the inferred power-law $g(L) \propto L^{-1.298}$ can explain the abundance of sources with large luminosities, the spatial density models are found to be practically indistinguishable from the current observations. In addition to the pulse broadening due to propagation effects, the observed flux distribution is also affected by the instrumental effects in the detection equipment such as the telescope beam shape and temporal resolution. While a coarse temporal resolution makes it less likely to detect a pulse with small $w_{\rm obs}$ due to the instrumental noise, there is also an observing bias against events with large $w_{\rm obs}$ due to reduced telescope sensitivity. 

We performed MC simulations to understand the effects of telescope observing biases, FRB energy density function and host galaxy properties on the observed flux distribution. We found that FRBs are unlikely to originate from relatively older stars with $\rho(z) \propto \rho_{\rm SMD}$ and should have a luminosity function that is steeper than the inferred $g(L) \propto L^{-1.298}$ based on the current detection rate of the brighter events with Parkes. Figure \ref{fig4} shows the comparison of observed Parkes $S_{\rm peak}$ with that from simulations for $\alpha=0$, $\beta=1.0$ and SFH $\rho(z)$ model. The simulations are carried out for $g(L) \propto {\rm exp}(-L/5.1\times10^{44}\ {\rm erg/s})$, $g(L) \propto L^{-1.298}$ and $g(L) \propto L^{-3.0}$. We find that the source luminosity function is better modelled with a relatively steeper power-law $g(L)$ with index $\lesssim -3.0$ or an exponential $g(L)$ with luminosity cutoff $L_{\rm c} \sim L_{\rm mid}$. 

Based on the current Parkes observations, we have found that the FRB progenitors are most likely to be younger stars with spatial density tracing the cosmic SFH, have a relatively flat source energy density spectrum with $\alpha \approx 0$ and a host galaxy DM contribution $\beta \approx 1$ that is similar to that from the MW. As the observed sample of FRBs further grows with detections made at finer temporal resolutions and with better source localisations across multiple surveys, stronger constraints can be applied using our analysis on the source luminosity function and the evolutionary history of the cosmic rate density from the observed flux distribution.




\label{lastpage}

\end{document}